\documentstyle[prl,aps,twocolumn,epsf]{revtex}
\begin{document}
\draft
\title{A Multilevel Blocking Approach 
to the Sign Problem in Real-Time Quantum Monte Carlo Simulations}
\author{C.H. Mak$^1$ and R.~Egger$^2$}
\address{${}^1$Department of Chemistry, 
University of Southern California, Los Angeles, CA 90089-0482\\
${}^2$Fakult\"at f\"ur Physik, Albert-Ludwigs-Universit\"at,  
 D-79104 Freiburg, Germany}
\date{Date: \today}
\maketitle
\begin{abstract}
We propose a novel approach toward the general solution of 
the sign problem in real-time path-integral simulations.
Using a recursive multilevel blocking strategy, this method
circumvents the sign problem by synthesizing the phase 
cancellations arising from interfering quantum paths 
on  multiple timescales.  
A number of numerical examples with one or a few explicit degrees 
of freedom illustrate the usefulness of the method.
\end{abstract}

\narrowtext

Path integrals \cite{feynman} provide an elegant 
alternative to the operator formulation of quantum mechanics.
Because they are easily adapted to many-body systems, 
quantum Monte Carlo (QMC) 
simulations based on path integrals can potentially yield
 exact results for the dynamics of condensed phase quantum systems.
A number of early attempts to use QMC simulations 
for real-time dynamics \cite{86thi401} demonstrated their potential, 
but these studies also uncovered the ubiquitous ``dynamical sign problem'' 
--- interference among quantum paths leads to large statistical noise
that increases linearly with the number of
possible paths, which in turn grows exponentially with the
timescale of the problem. 
Consequently, real-time QMC simulations have been limited 
to problems of very short timescales.
Several attempts to extend the timescale of real-time QMC simulations
have appeared \cite{86fil717,87mak10,88dol277}, all of which 
rely on a common idea ---
by using various filtering schemes, the
phase cancellations can be numerically stabilized by damping out the
non-stationary regions in configuration space.  
Such filtering methods were able to extend the 
timescale somewhat, but they were generally unable 
to reach timescales of interest in typical chemical systems.

Here we propose a novel approach based on a
{\em blocking strategy} which may provide a general solution 
to the dynamical sign problem for very long times. 
The blocking strategy asserts that by sampling
groups of paths called {\em blocks},
the sign problem is {\em always} reduced compared to 
sampling single paths \cite{96mak39}.
Though this blocking strategy seems simple, its practical
implementation is cumbersome, especially when going out to long 
times.  Because the number of paths grows exponentially 
with the timescale, the number of blocks also grows immensely.
To cure this problem, we first define elementary blocks 
and then group them together into larger blocks.  
Blocks of different sizes are introduced
on several timescales called {\em levels}.  
After taking care of the sign cancellations
within all blocks on a finer level, the entire sign 
problem can then be transferred to the next coarser level.
By recursively proceeding from the bottom level (shortest timescale) 
up to the top (longest timescale), the troublesome numerical
instabilities associated with the sign problem can be
systematically avoided.  In a slightly different form, this
{\em multilevel blocking}  (MLB) algorithm has
recently been applied to treat the 
``fermion sign problem'' in many-fermion 
 imaginary-time simulations \cite{own}. 
As most technical details of the MLB scheme can be found in \cite{own},
we only give a brief description of the algorithm below,
stressing the main ideas and its differences from the 
fermion formulation.

For concreteness, we describe the MLB method for the calculation of
an equilibrium time-correlation function, 
\begin{equation}\label{corr}
\langle A(0) B(t) \rangle = 
\frac{{\rm Tr} \left\{e^{-(\beta\hbar+it)H/\hbar} A e^{+itH/\hbar} B \right\}}
     {{\rm Tr} \left\{
e^{-(\beta\hbar+it)H/\hbar}   e^{+itH/\hbar} \right\}  }\;.
\end{equation}
With minor modifications, the MLB method also applies to other 
dynamical properties like the thermally symmetrized correlation function 
\cite{84thi5029},
$C_s(t) = Z^{-1} {\rm Tr}\{e^{-(\beta\hbar/2+it)H/\hbar} 
A e^{-(\beta/\hbar/2-it)H/\hbar} B\}$, with $Z$ being the partition 
function. 
In terms of path integrals, the traces in (\ref{corr}) involve 
two quantum paths, one propagated backward in time for the duration
$-t$ and the other propagated in complex time for the duration 
$t-i\beta\hbar$. 
Discretizing each of the two paths into $P$ slices, the entire
cyclic path has a total of $2P$ slices. A slice on the first half of them
has length $-t/P$, and on the second half $(t-i\beta\hbar)/P$.  
We require $P=2^L$ which defines the total number of levels $L$.
Denoting the quantum numbers (e.g., spin or position variables)
at slice $j$ by $\bbox{r}_j$, 
$\{\bbox{r}_1, \cdots \bbox{r}_{2P}\}$ is a discrete representation 
of a path, and 
the correlation function (\ref{corr}) reads
\begin{equation}\label{naive}
\frac{ \int d\bbox{r}_1 \cdots d\bbox{r}_{2P}
       B(\bbox{r}_{2P}) A(\bbox{r}_{P}) 
       \prod_{j=1}^{2P} (\bbox{r}_j,\bbox{r}_{j+1})_0 }
     { \int d\bbox{r}_1 \cdots d\bbox{r}_{2P}
       \prod_{j=1}^{2P} (\bbox{r}_j,\bbox{r}_{j+1})_0 }\;,
\end{equation}
where the {\em level-0 bond}
$(\bbox{r}_j,\bbox{r}_{j+1})_0$ is simply the
short-time propagator between slices $j$ and $j+1$, 
and $\bbox{r}_{2P+1}= \bbox{r}_1$.  
A direct application of the QMC method would sample these paths
 using the modulus of the 
integrand in the denominator as the weight. 

\begin{figure}
\epsfxsize=0.8\columnwidth
\epsffile{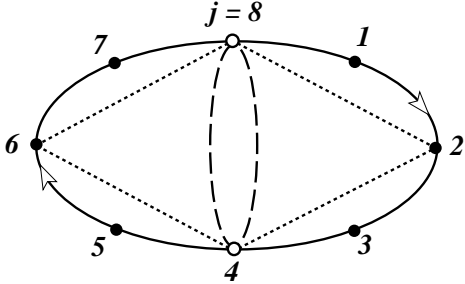}
\caption[]{\label{fig1}
Example of how slices (circles) are assigned to various 
levels for $L=2$. 
The first path goes from $j=8$ to $4$ and the second from $j=4$ to $8$.
The finest level $\ell=0$
contains  $j=1,3,5,7$, level $\ell=1$ contains $j=2,6$, and
the top level $\ell=2$ contains $j=4,8$. 
Coarse bonds are indicated by dotted (level-1) and dashed (level-2)
lines.
}
\end{figure}

We first assign all slices along the discretized path
to different levels $\ell=0,\ldots,L$ 
(see Figure \ref{fig1}).  
Each slice $j=1,\ldots,2P$ belongs to a unique level $\ell$, 
such that $j = (2k+1)2^\ell$
and $k$ is a nonnegative integer.  
For instance, slices $j = 1, 3, 5, \ldots$ belong to level $\ell = 0$, 
slices $j = 2, 6, 10, \ldots$ to $\ell = 1$, etc.
The MLB algorithm starts by sampling only configurations
which are allowed to vary on slices associated with the
finest level $\ell = 0$,  using the weight
${\cal P}_0=|(\bbox{r}_1,\bbox{r}_2)_0 \cdots (\bbox{r}_{2P},\bbox{r}_1)_0|$.
The short-time level-$0$ bonds are then employed to
 synthesize longer-time level-$1$
bonds that connect the even-$j$ slices.
Subsequently the level-1 bonds are used
to synthesize level-$2$ bonds, and so on.
In this way the MLB algorithm moves recursively from 
the finest level ($\ell = 0$) up to increasingly coarser levels 
until $\ell = L$, where the measurement
is done using $\bbox{r}_{2P}$ and $\bbox{r}_{P}$.

Generating a MC trajectory containing $K$ samples for each slice 
on level $\ell=0$ and storing these samples, 
we compute the {\em level-1 bonds} according to
\begin{eqnarray} \label{l1b}
(\bbox{r}_j,\bbox{r}_{j+2})_1 &=&  \langle 
{\Phi} [ (\bbox{r}_j,\bbox{r}_{j+1})_0 (\bbox{r}_{j+1},\bbox{r}_{j+2})_0 ]
 \rangle_{{\cal P}_0} \\ \nonumber
&=& K^{-1} \sum_{\bbox{r}_{j+1}} {\Phi} [ (\bbox{r}_j,\bbox{r}_{j+1})_0 
(\bbox{r}_{j+1},\bbox{r}_{j+2})_0] \;,
\end{eqnarray}
where the summation $\sum_{\bbox{r}_{j+1}}$ extends over the $K$ samples,
and $\Phi[z] = e^{i{\rm arg}(z)}$ denotes the phase.
For a complete solution of the sign problem, 
the sample number $K$ has to be sufficiently large \cite{own}.
The bonds (\ref{l1b}) contain crucial information about the
sign cancellations on the previous level $\ell=0$.
Their benefit becomes clear when rewriting
the integrand of the denominator in (\ref{naive}) as
\[
{\cal P}_0 \, \times \, (\bbox{r}_2,\bbox{r}_4)_1 \cdots 
(\bbox{r}_{2P-2},\bbox{r}_{2P})_1 (\bbox{r}_{2P},\bbox{r}_2)_1 \;.
\]
Comparing this to (\ref{naive}), we notice that 
the entire sign problem has been transferred to the next coarser level.

In the next step, the sampling is carried
out on level $\ell=1$ in order to compute the next-level bonds, 
using the weight ${\cal P}_0 {\cal P}_1$ with 
${\cal P}_1 = |(\bbox{r}_2,\bbox{r}_4)_1  \cdots (\bbox{r}_{2P},\bbox{r}_2)_1|$.
 Generating a sequence of $K$ samples for each slice on level
 $\ell=1$, and storing these samples, 
we then calculate the {\em level-2 bonds} 
in analogy with (\ref{l1b}),
\[
(\bbox{r}_j,\bbox{r}_{j+4})_2 =  
\langle {\Phi}\left[ (\bbox{r}_j,\bbox{r}_{j+2})_1
(\bbox{r}_{j+2},\bbox{r}_{j+4})_1 \right]
 \rangle_{{\cal P}_0 {\cal P}_1 } \;,
\]
and iterate the process up to the top level
by employing analogously defined {\em level-$\ell$ bonds}.
The correlation function (\ref{corr}) can then be computed from 
\begin{equation}\label{expec}
\frac{\langle B(\bbox{r}_{2P}) A(\bbox{r}_P) 
 {\Phi}[(\bbox{r}_P,\bbox{r}_{2P})_L (\bbox{r}_{2P},\bbox{r}_P)_L]
\rangle_{\cal P}}
     {\langle{\Phi}[(\bbox{r}_P,\bbox{r}_{2P})_L 
(\bbox{r}_{2P},\bbox{r}_P)_L] \rangle_{\cal P}} \;,
\end{equation}
with the positive definite MC weight 
 ${\cal P} = {\cal P}_0 {\cal P}_1 \cdots {\cal P}_L$.
The denominator in (\ref{expec}) gives the {\em average phase} 
and indicates to what extent the sign problem has been solved. 
Under the direct QMC method, the average phase decays
exponentially with $t$ and is typically close to zero. 
With the MLB algorithm, however, the average phase remains close to unity 
even for long times, with a CPU time requirement 
$\sim t$. The price to pay for the stability of the
algorithm is the increased memory requirement $\sim K^2$ 
associated with having to store the sampled configurations.

Now we illustrate the practical usefulness of the MLB method by
several numerical examples.  In each of these examples, we 
compute a time-correlation function. 
The average phase is larger than 0.6 for all data sets shown below.
 The decay in the average phase with $t$ is a result of 
the finiteness of $K$ \cite{own}.  Choosing a
larger $K$ allows for a larger average phase out to
longer time at the cost of increased computer memory and CPU time.
Each data point in even the most intensive calculation took no more than a 
few hours on an IBM RS 6000/590.

\begin{figure}
\epsfxsize=0.8\columnwidth
\epsffile{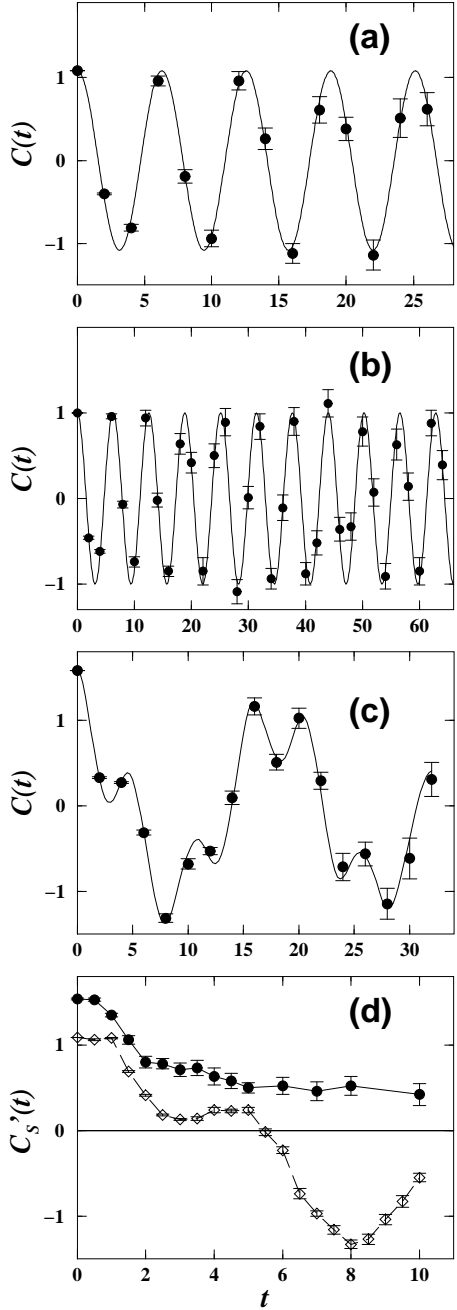}
\caption[]{\label{fig2}
MLB results (closed circles) for various systems.
Error bars indicate one standard deviation. 
(a) $C(t)$ for a harmonic oscillator at $\beta\hbar = 1$. 
The exact result is indicated by the solid curve. 
(b) Same as (a) for a two-level system at $\beta\hbar=10$.
(c) Same as (a) for a double-well system at $\beta\hbar=1$.
This temperature corresponds to the classical barrier energy.
(d) $C'_s(t)$ for a double-well system coupled to two
 oscillators at $\beta\hbar = 1$.
For comparison, open diamonds are for the uncoupled $(\alpha=0)$ system.
Note that $C'_s(t)$ is similar but not identical to $C(t)$ shown
in (c). Solid and dashed lines are guides to the eye only.}
\end{figure}

{\em A. Harmonic oscillator}.  
For $H = p^2/2m + m\omega^2 x^2/2$, 
the real and imaginary parts of 
$\langle x(0)x(t) \rangle$ oscillate in time due to vibrational coherence. 
In dimensionless units $m=\omega=1$,
the oscillation period is $2\pi$.
Figure \ref{fig2}(a) shows MLB results
for $C(t)={\rm Re}\, \langle x(0) x(t) \rangle$.
With $P=32$ for the maximum time $t=26$, 
$K = 200$ samples were used for sampling the coarser bonds.
Within error bars, the data coincide with the exact result 
and the algorithm produces stable results free of the sign problem. 
Without MLB, the signal-to-noise ratio was practically zero 
for $ t > 2$.

{\em B. Two-level system}.
For a symmetric two-state system, 
$H = -\frac{1}{2} \Delta \sigma_x$, 
 the dynamics is controlled by tunneling.
The spin correlation function $\langle \sigma_z(0)\sigma_z(t) \rangle$ 
exhibits oscillations indicative of quantum coherence.  
Figure \ref{fig2}(b) shows MLB results for
 $C(t)={\rm Re}\,\langle \sigma_z(0)\sigma_z(t) \rangle$,
Putting $\Delta=1$, the tunneling oscillations have a period of $2\pi$.
With $P = 64$ for the maximum time $t=64$,
only $K = 100$ samples were used for sampling the coarser bonds.
The data agree well with the exact result.  
Again the simulation is stable and free of the sign problem. 
Without MBL, the simulation failed for $t > 4$.

{\em C. Double-well potential}.
Next, we examine a double-well system 
with the quartic potential $V(x) = -x^2 + \frac{1}{4}x^4$.
At low temperatures, interwell transfer occurs through tunneling motions 
on top of intrawell vibrations.  These two effects
combine to produce nontrivial structures in the position correlation
function.  At high temperatures,
interwell transfer can also occur by classical barrier crossings.
Figure \ref{fig2}(c) shows MLB results
for $C(t)={\rm Re}\,\langle x(0)x(t)\rangle$.
The slow oscillation corresponds to interwell tunneling,
with a period of approximately 16.  The higher-frequency 
motions are characteristic of intrawell oscillations.  
In this simulation, $K = 300$ samples were used.
The data reproduce the exact result well, capturing
all the fine features of the oscillations.  
Again the calculation is stable and free of the sign problem,
whereas a direct simulation failed for $t>3$.

{\em D. Multidimensional tunneling system}.
As a final example, we consider a problem with three degrees of freedom, 
in which a particle in a double-well potential is bilinearly coupled to 
two harmonic oscillators.
The quartic potential in the last example is used for the double-well, 
and the harmonic potential in the first example is used for both oscillators.
The coupling constant between each oscillator and the tunneling particle 
is $\alpha = 1/2$ in dimensionless units.
For this example, we computed the correlation function $C_s(t)$ 
for the position operator of the tunneling particle.
Direct application of MC sampling to $C_s(t)$
has generally been found unstable for $t>\beta\hbar/2$
\cite{84thi5029}.
In contrast, employing only moderate values of $K=400$ to 900, 
the MLB calculations allowed us to go up 
to $t=10\beta\hbar$.  
(Notice that this $K$ is no larger than three, i.e. the 
number of dimensions, times the $K$ needed for one-dimensional 
systems.)
Figure \ref{fig2}(d) shows MLB results
for $C'_s(t)={\rm Re} \,C_s(t)$.
For the coupled system, the position correlations have lost the coherent 
oscillations and instead decay monotonically with time. 
Coupling to the medium clearly damps the coherence and tends to
 localize the tunneling particle.

The data presented here demonstrate that 
the MLB method holds substantial promise toward an
 exact and stable simulation method for 
real-time quantum dynamics computations of many-dimensional systems up
to timescales of practical interest. 
Instead of the exponentially vanishing signal-to-noise ratio in a 
ordinary application of the Monte Carlo method to real-time path 
integral problems, the MLB method has a CPU requirement that grows 
only linearly with time.  Moreover, the data we have so far seem to 
suggest that the memory requirement $K$ also grows only linearly 
with the dimensionality of the system.

This research has been supported by the National Science Foundation
under grants CHE-9257094 and CHE-9528121, by the Sloan Foundation,
the Dreyfus Foundation, and by the Volkswagen-Stiftung.

\end{document}